\documentclass[a4paper]{article}
\usepackage{hyperref}
\usepackage{INTERSPEECH2020}
\usepackage{multirow}
\usepackage{subfigure}
\usepackage[symbol]{footmisc}
\usepackage{enumitem}
\newlist{steps}{enumerate}{1}
\setlist[steps, 1]{label = Step \arabic*:}
\usepackage[ruled,vlined]{algorithm2e}
\usepackage[compress]{cite}

\title{The NeteaseGames System for Voice Conversion Challenge 2020 with Vector-quantization Variational Autoencoder and WaveNet}
\name{ Haitong         Zhang}
\address{
 NetEase Games AI Lab
  }
\email{zhanghaitong01@corp.netease.com}

\begin{document}

\maketitle

\begin{abstract}
This paper presents the description of our submitted system for Voice Conversion Challenge (VCC) 2020 with vector-quantization variational autoencoder (VQ-VAE) with WaveNet as the decoder, i.e., VQ-VAE-WaveNet. VQ-VAE-WaveNet is a nonparallel VAE-based voice conversion that reconstructs the acoustic features along with separating the linguistic information with speaker identity. The model is further improved with the WaveNet cycle as the decoder to generate the high-quality speech waveform, since WaveNet, as an autoregressive neural vocoder, has achieved the SoTA result of waveform generation. In practice, our system can be developed with VCC 2020 dataset for both Task 1 (intra-lingual) and Task 2 (cross-lingual). However, we only submit our system for the intra-lingual voice conversion task. The results of VCC 2020 demonstrate that our system VQ-VAE-WaveNet achieves: 3.04 mean opinion score (MOS) in naturalness and 3.28 average score in similarity ( the speaker similarity percentage (Sim) of 75.99\%) for Task 1. The subjective evaluations also reveal that our system gives a top performance when no supervised learning is involved. What's more, our system performs well in some objective evaluations. Specifically, our system achieves an average score of 3.95 in naturalness in automatic naturalness prediction and ranked the 6th and 8th, respectively in ASV-based speaker similarity and spoofing countermeasures.

\end{abstract}

\section{Introduction}

Voice conversion (VC) aims to changing the speaker identity of a source utterance into that of a desired target speaker while retaining the linguistic contents of the source utterance \cite{abe1990voice}. VC is very useful for various new applications, such as expressive speech synthesis \cite{kain1998spectral}, impaired speech improvement \cite{tanaka2013hybrid}. 

Voice conversion challenge (VCC) \footnote{ \url{http://vc-challenge.org} } aims to better understand different VC techniques built on a common released dataset. During the past six years, there VC challenges have been carried out, i.e., VCC 2016 \cite{toda2016voice}, VCC 2018 \cite{lorenzo2018voice}, and VCC 2020 \cite{vcc2020}, with a common goal, but a slight different focus. In VCC 2020, two new tasks are considered. The first one is semi-parallel intra-lingual VC, where the training data comes from one language and small part of them are parallel. The second task is cross-lingual VC, where the source speaker is different from the target one both in terms of language and content. 

Since the challenge focus on non-parallel VC, it should be noted that a promising paradigm for non-parallel VC is a recognition-synthesis framework \cite{sun2016phonetic, liu2018wavenet}. The idea is to first extract the linguistic content from the source speech and concatenated with the speaker identity/feature to generate the speech in target voice. A popular type of models are based on phonetic posteriorgram (PPG). In this type of models, the source speech is usually fed into an automatic speech recognition (ASR) model to extract the PPG, then a synthesis model is applied to synthesize the speech in target speaker's voice by conditioning on the PPG. Another group of VC techniques for non-parallel VC are based on auto-encoder-like model \cite{hsu2016voice, kameoka2018acvae, qian2019autovc}. These models are developed to disentangle the speaker characteristics from the linguistic contexts by using only reconstruction on the spectral feature level. 

Both types of techniques mentioned have been proved efficient in non-parallel VC. However, considering the first type of techniques reply much on the performance ASR model (which also requires much more work on data annotation/labelling), we focus on using auto-encoder-like model to perform non-parallel VC for Task 1. 

Specifically, we built our system based on VQ-VAE model \cite{van2017neural}. We used VQ-VAE mainly because the vector-quantization part can discretize the continuous representations into a fixed-number of embeddings, and these embeddings are proven to be phoneme-like in \cite{chorowski2019unsupervised, zhang2020unsupervised}. Thus, we believe VQ-VAE would perform better in extracting the linguistic context. 

Many great VC systems consist of a conversion model or a neural vocoder \cite{oord2016wavenet, kalchbrenner2018efficient}, with the conversion model to convert the source acoustic features into target speaker's voice, and the vocoder to transform the converted features into speech waveform. These two models are usually trained in a separate mode. As a result, there would be some mismatches between the data distribution of the converted features and that of the ground truth features used to train the vocoder. Usually these mismatches could be reduced by training or fine-tuning the neural vocoder by using the converted features and/or the natural features. This can be regarded as a method of data augmentation to regularize the neural vocoder. Based on this observation, we combined the vocoder part into the conversion model in order to train the models jointly, which can perform voice conversion and produce the speech waveform directly.

The rest of this paper is organized as follows: Section 2 details our system. Section 3 provides the evaluation results and our system performance. Conclusions are given in Section 4.

\section{NeteaseGames System} \label{proposed}

\begin{figure}[t]
  \centering
  \includegraphics[ width=8cm, height=12cm]{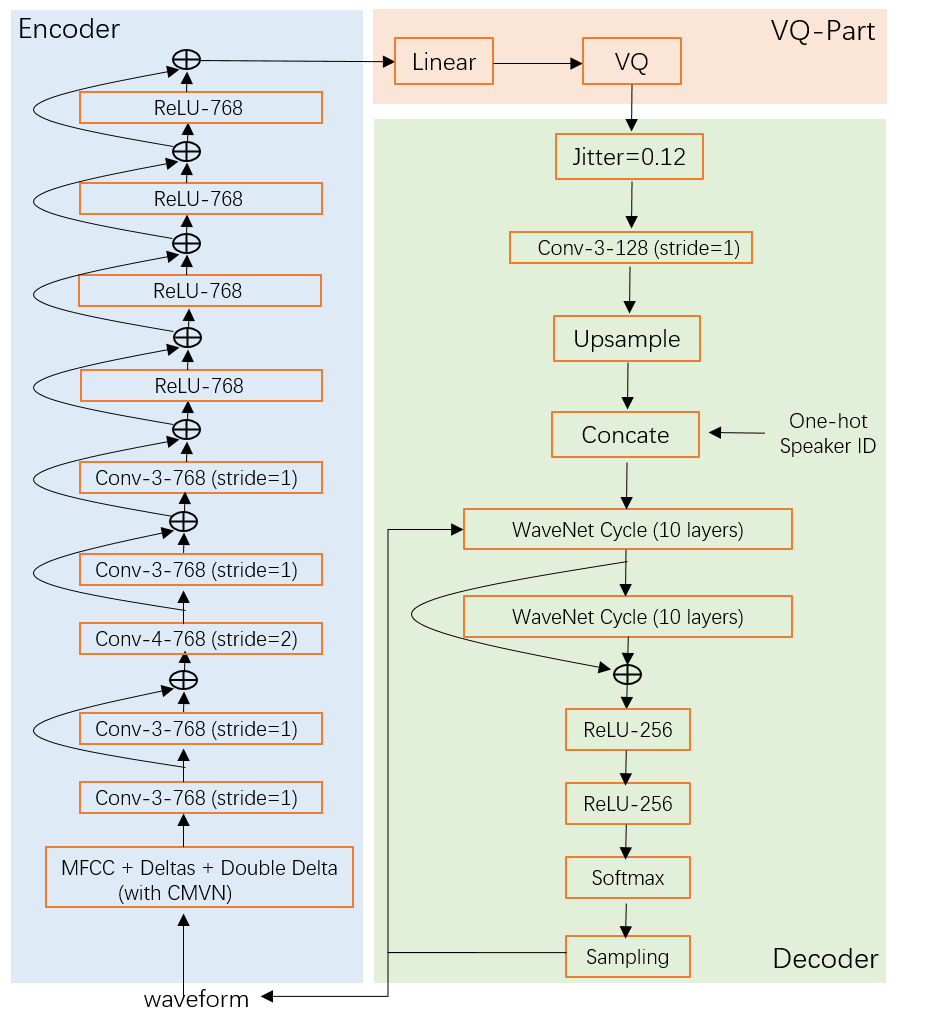}
  \caption{The model structure of our VC system.}
  \label{fig:model}
\end{figure}

In this section, we describes the overall structure of our submitted system for Task 1 of VCC 2020, as illustrated in Figure \ref{fig:model}. In details, our system mainly consists of three parts, namely the encoder, the vector-quantization part, WaveNet-based decoder. We describe each module as follows.

\subsection{Encoder}

The encoder part is partly composed of stacks of convolutional layers and fully-connected layers, and residual connections. The encoder takes the source features $X={x_1, x_2, ... ,x_T}$ and learn the high-level representation through non-linear transformation. The whole procedure is formulated as below.

\begin{equation}
    \begin{aligned}
    h_1 &= Conv_{3 * 768}(X); \\
    h_2 &= Conv_{3 * 768}(h_1) + h_1;  \\
    h_3 &= Conv_{4 * 768}(h_2), stride = 2 \\
    h_4 &= Conv_{3 * 768}(h_3) +h_3;  \\
    h_5 &= Conv_{3 * 768}(h_4) + h_4; \\
    h_6 &= ReLU( w h_5 + b) + h_5; \\
    h_7 &= ReLU( w h_6 + b) + h_6; \\
    h_8 &= ReLU( w h_7 + b) + h_7; \\
    h_9 &= ReLU( w h_8 + b) + h_8
    \end{aligned}
\end{equation}

where $3*768$ means the kernel size is $3$ and filter size is $768$,  $+$ means residual connection, $w$ and $b$ is the weight matrix and bias. Downsampling effect is achieved by setting stride step into $2$ in the third convolutional layer.

\begin{figure}[t]
  \centering
  \includegraphics[ width=8cm]{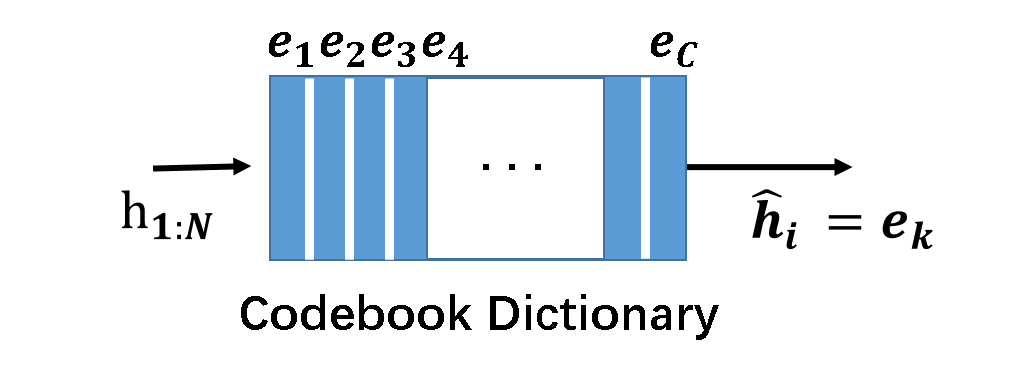}
  \caption{The VQ part in VQ-VAE.}
  \label{fig:VQ}
\end{figure}

\subsection{Vector quantization Part}

Vector quantization (VQ) part is an important part in VQ-VAE, used to quantization the continuous hidden representation into fixed number of embeddings, as illustrated in Figure \ref{fig:VQ}. 

A codebook dictionary of size $e = C*D$ is pre-defined, where $C$ is the number of latent embeddings in the dictionary and $D$ is the dimension of each embedding. The VQ part takes the encoded representation $h_9 = h_{1:N}$ , where $N$ depends on the length $T$ and the number of down-sampling layers in the encoder (i.e. N =  T / down-sampling factor). Then the continuous latent representations $h_{1:N}$ can be mapped into $\hat{h}_{1:N}$ by finding the nearest pre-defined discretized embedding in the dictionary as  $ \hat{h} = e_k$ , where $k = argmin_j$  $|| h - e_j||$, and $e_j$ is the j-th embedding in the codebook dictionary, and $j \in  {1,2, \cdots ,C}$.

\subsection{Jitter Mechanism}

Jitter mechanism is inspired by \cite{chorowski2019unsupervised}. We insert the jitter mechanism after the VQ part in order to enforce some consecutive frames would be quantized into the same embedding in the codebook dictionary, since the speech signal changes slowly.  The jitter mechanism replace the embedding at current frame with that at the last frame and at the next frame at a certain probability.

\subsection{Decoder}

After jitter mechanism, the learned embedding sequence is further processed by one convolutional layer before upsampling layer. Unsampling is used to match the length of the hidden representation sequence and the input feature sequence length. After upsampling, the sequence is concatenated with one-hot speaker embedding at each time index to form the final input of the WaveNet-like decoder.

\begin{figure}[t]
  \centering
  \includegraphics[ height=8cm]{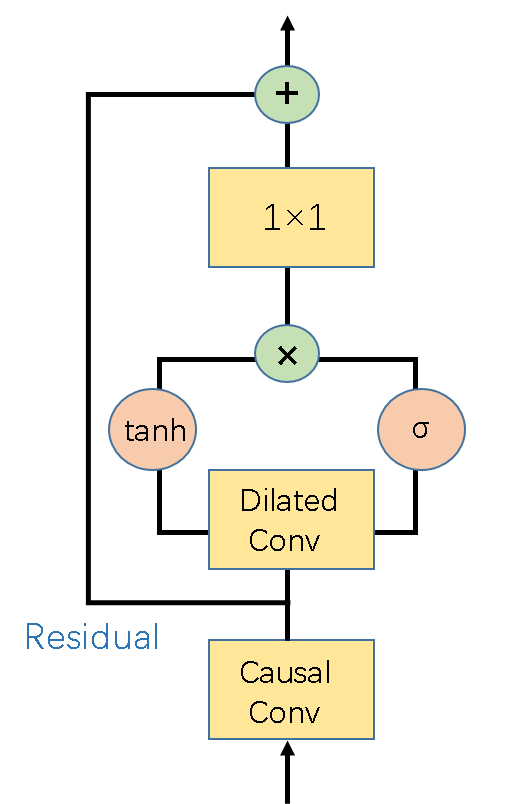}
  \caption{The main structure in WaveNet-like decoder.}
  \label{fig:wavenet}
\end{figure}

Conventional vocoders based on source-filter theory severely limit the synthesis quality \cite{morise2016world, kawahara2006straight} . The neural-network-based vocoders greatly improve the quality of synthesized speech by directly modeling the speech waveform and bypassing the traditional speech vocoders. WaveNet-based neural vocoder, especially, could generate the waveform whose quality is comparable to that of the natural speech. WaveNet is an autoregressive generative model that models waveform directly. The original WaveNet is realized by the use of conditional WaveNet, in which acoustic feature is used as the conditional input to generate waveform (see Figure \ref{fig:wavenet}). Given a sequence of waveform $X = {x_1, x_2, ..., x_t}$, the joint probability of all these samples is represented as follows:

\begin{equation}
    p(x| h; \theta) = \sum_{t=1}^{T} p(x_t| x_1, x_2,...,x_{t-1}, h; \theta)
\end{equation}

where $h$ is the local conditional features, $\theta$ represents the parameters of the WaveNet part. $p(x_t| x_1, x_2, ...,x_{t-1})$ denotes a conditional dependence between current waveform samples and the previous samples in a long range. In original WaveNet model, the local conditional features are ground truth acoustic features, in our model, these local conditional features are the learned features given the ground truth features through the mentioned nonlinear transformation.

As the WaveNet architecture in \cite{oord2016wavenet}, we adopted 20 dilated convolution layers, grouped into 2 dilation cycles, i.e., the dilation rate of layer $k(k = 0, 1, ..., 19)$ is $2^{(k(mod10))}$. The filter width is 2, and residual and skip channels are set into 256, and the softmax output dimension is 128. The model is optimized with Adam algorithm.

\subsection{Training Setup}

The training data released for Task 1 of Voice Conversion Challenge 2020 consists of 8 speakers in English language. That of source speakers consist of 2 males and 2 females. The set of target speakers includes another 4 English speaker (2 males and 2 females). The number of utterances for each speaker is 70 with 20 parallel utterances between source and target speakers. To further improve the model robustness to unseen data, we augmented the training data with speech of 100 speakers from VCTK speech corpus \cite{veaux2016superseded}. The model was trained with a mixture of training data released and VCTK data, then fine-tuned with the training data released only. We kept five utterances from each speaker for validation purpose. We did not perform any data labelling and checking. 

We used MFCC as model inputs instead of waveform, since we believe using MFCC is beneficial to disentangling speaker characteristics from the input features. Specifically, we used first 13 MFCC features. We then concatenate MFCC with delta and double delta features and then apply CMVN across each speaker. We set the jitter probability into $0.12$ by empirical experiments. The whole model is trained using Adam algorithm, with a starting learning rate of $2.5e^{-4}$ and halving into $1e^{-5}$ until convergence.

\section{Evaluations}

Three baseline systems are also included in the task. The first baseline system was built using PPG-based VC techniques, which achieved the best performance in VCC 2018 \cite{liu2018wavenet}, but not publicly available. The second baseline is a cascaded ASR and TTS system using sequence-to-sequence models \cite{Huang2020}, whose open implementation is freely available at  \url{https://github.com/espnet/ espnet/tree/master/egs/vcc20}. The third baseline is CycleVAEPWG, where CycleVAE is used for voice conversion and Parallel WaveGAN (PWG) for waveform generation \cite{Tobing2020}, and the implementation has been made freely available at  \url{https://github.com/bigpon/vcc20_baseline_cyclevae}. Including three baselines, there were 31 participant systems submitted to Task 1. 

The official evaluation test consists of 25 unseen utterances in English. In the task one, each source utterances are required to be converted into four target speakers' voice. Thus, the whole set of submitted utterances for the task includes in total 400 utterances. But only 5 utterances (E30001, E30002, E30003, E30004, and E30005) are selected for perceptual evaluation. A large-scale listening test was conducted for subjective evaluation, where there were 206 Japanese listeners and 68 English listeners.

\subsection{Naturalness Test}

\begin{figure}[t]
  \centering
  \includegraphics[ width=8cm]{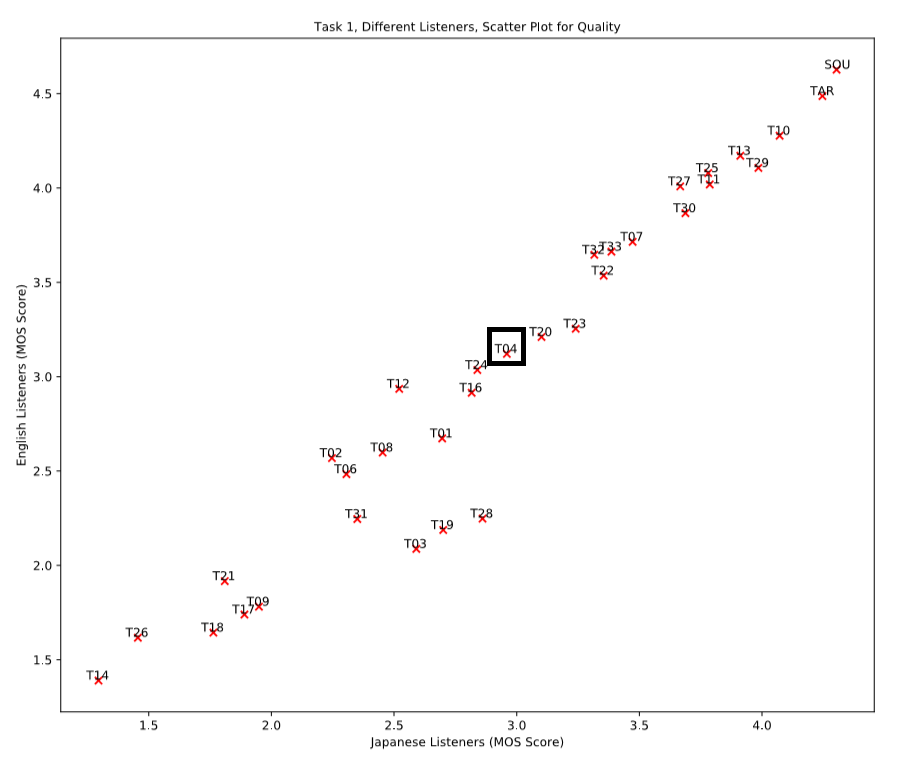}
  \caption{Scatter plot matching naturalness of Japanese listeners and English listeners for Task 1.}
  \label{fig:natural}
\end{figure}

Figure \ref{fig:natural} shows the scatter plot matching naturalness of Japanese listeners and English ones. Three baselines are denoted as T11, T22, and T16, respectively. TAR and SOU refer to the natural speech of the target speaker and the source speaker, respectively. From the plot, we can witness that four systems perform better than T11, the best performing system in VCC 2018, and the difference between T10 or T13 and T11 is significant. It indicates that VC performance has improved within the past two years in terms of naturalness. Another important highlight is that most of the top performing systems are based on PPG or ASR or leveraged with TTS, which indicates that more labelled data is helpful for VC performance. Our system, T04, achieved an average MOS score of 3.04 (the average score of Japanese listeners and English ones). It should be noted that another two systems based on VQ-VAE achieved an average score of 2.44 and 3.16. Although the PPG or recognition-based models achieve the best performances, our systems, together with T20, achieved the best performance in naturalness as no supervised learning (such as ASR or TTS) is involved.

\subsection{Similarity Test}
\begin{figure}[t]
  \centering
  \includegraphics[ width=8cm]{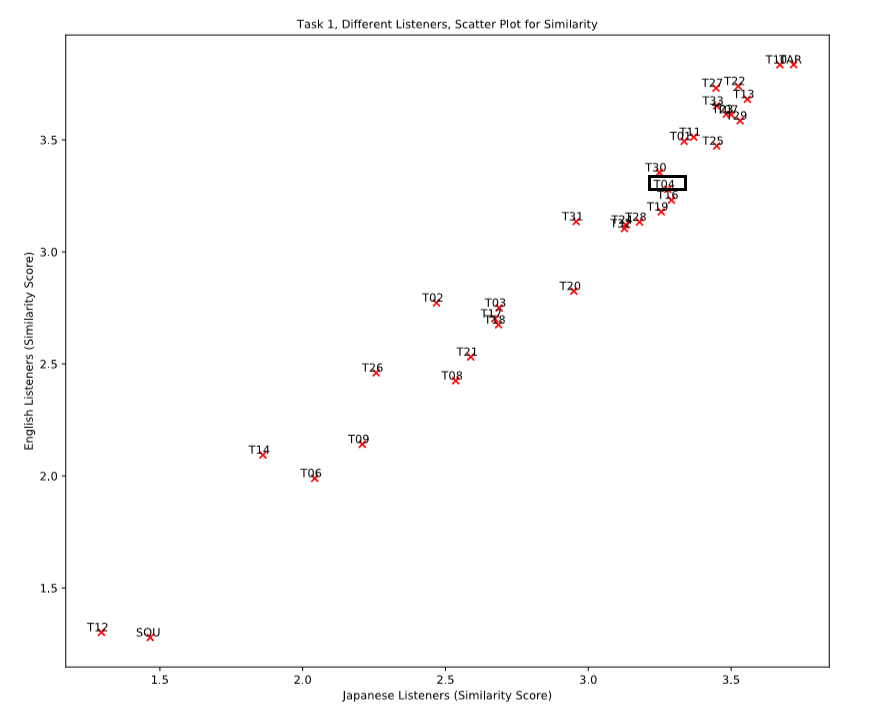}
  \caption{Scatter plot matching similarity of Japanese listeners and English listeners for Task 1.}
  \label{fig:similar}
\end{figure}

Figure \ref{fig:similar} shows the scatter plot matching similarity of Japanese listeners and English ones. It is shown that our system achieved an average score of 3.28 in similarity. What is interesting is that system T20 achieved an average score of 2.89. This difference could be attributed to the joint training strategy of our system. Our system also perform the best as no unsupervised learning is considered.

\begin{table*}
\begin{center}
\caption{Objective assessments on VCC 2020, which is adapted from \cite{vcc2020_ObjEval}.}
\begin{tabular}{ |c|c|c|c| } 
\hline
Measurement Tool & Type of Measure & Metric \\
\hline
Automatic Speaker Verification &  Target-Spoof Assessment & Equal Error Rate \\ 
Automatic Speaker Verification &  Converted Source $\Longleftrightarrow$ Target Similarity & Pfa(tar) \\ 
Automatic Speaker Verification &  Converted Source $\Longleftrightarrow$ Source Similarity & Pmiss(src)     \\
Spoofing Countermeasure        &  Real-Fake Assessment  & Equal Error Rate  \\
Objective Mean Opinion Score   &  Quality   &  Mean Opinion Score \\
Automatic Speech Recognition   &  Intelligibility   & Word Error Rate \\
\hline
\end{tabular}
\label{table:OBJ}
\end{center}
\end{table*}

\subsection{Objective evaluation}

Besides the subjective evaluation results, objective evaluations were also conducted \cite{vcc2020_ObjEval}. The evaluations include (1) text-independent ASV for speaker similarity, text-independent CM for real-vs-fake assessment, automatic MOS prediction for quality, and ASR for intelligibility, as shown in Table \ref{table:OBJ}.

 Three kinds of ASV errors are included in the assessment, namely ASV EER, $P_{fa}^{tar}$, and $P_{miss}^{src}$. Overall, our system T04 ranked the 6th, achieving $45.13\%$ , 99\% , and 100\% respectively for these three criteria.

The spoofing countermeasures were also included. The spoofing countermeasures is useful to defending various kinds of attacks to ASV systems. Our system ranked the 8th in all submitted systems.

An automatic predicted MOS \cite{lo2019mosnet} was also reported. Our system achieved the 3rd place among all systems, with an average score of 3.92. But the predicted MOS decreases into 3.25 as a different training dataset is used for the prediction model. At last, our system achieved a WER of about 20\%.

\section{Conclusion}

This paper presents the details of our submitted system and summarizes the results in VCC 2020. We built our system based on VQ-VAE model. To further improve the performance we replaced the decoder in VQ-VAE with WaveNet-like decoder to directly produce speech waveform. As a result, the whole model can be trained jointly. Jitter mechanism was also applied to improve disentangling the linguistic content and speaker characteristics by our preliminary experiments. The official results show that our model achieved a naturalness and similarity score above 3. And our system is the best performing one as far as no supervised learning is involved. What is interesting is that our system achieved the third place in automatic naturalness prediction. What's more, our model performed well in ASV based speaker similarity assessments and spoofing countermeasures, ranking the 6th and 8th, respectively.

Although our system achieved reasonably good results in the task, there is still a gap between our system performance and the top-performance systems based on recognition or PPG or TTS. It shows that incorporating more labelled data or supervised learning results in a better conversion performance.


\bibliographystyle{IEEEtran}
\bibliography{dataEfficiency}

\end{document}